\documentclass[useAMS,usenatbib]{mn2e}

\usepackage{times}
\usepackage{graphics,epsfig}
\usepackage{graphicx}
\usepackage{amsmath}

\newcommand{\kms}{km~s$^{-1}$}

\newcommand{\phip}{$\phi(p)$}
\newcommand{\psiq}{$\psi(q)$}

\title[Axial ratio of faint dwarfs]{Thick gas discs in faint dwarf galaxies}
\author[Roychowdhury et al.]{Sambit Roychowdhury$^{1}$\thanks{E-mail: sambit@ncra.tifr.res.in}, 
         Jayaram N. Chengalur$^{1}$\thanks{~~~~~~~~~~~~~chengalu@ncra.tifr.res.in}, 
         Ayesha Begum$^{2}$\thanks{~~~~~~~~~~~~~begum@astro.wisc.edu}
         \newauthor and Igor D. Karachentsev$^{3}$\thanks{~~~~~~~~~~~~~ikar@sao.ru}\\
       \\ 
       $^{1}$NCRA-TIFR, Post Bag 3, Ganeshkhind, Pune 411 007, India\\
       $^{2}$Dept of Astronomy, University of Wisconsin-Madison, Madison WI 53706-1582\\
       $^{3}$Special Astrophysical Observatory, Russian Academy of Sciences, N. Arkhyz, KChR 369167, Russia}

\begin{document}
\date{}

\pagerange{\pageref{firstpage}--\pageref{lastpage}} \pubyear{}

\maketitle

\label{firstpage}

\begin{abstract}

  We determine the intrinsic axial ratio distribution of the {\it gas} discs of extremely
faint M$_B > -14.5$ dwarf irregular galaxies. We start with the measured (beam corrected)
distribution of apparent axial ratios in the HI 21cm images of dwarf irregular galaxies
observed as part of the Faint Irregular Galaxy GMRT Survey (FIGGS). Assuming that the
discs can be approximated as oblate spheroids, the intrinsic axial ratio distribution
can be obtained from the observed apparent axial ratio distribution. We use a couple of methods to do this, and our final results are based on using Lucy's deconvolution
algorithm. This method is constrained to produce physically plausible distributions,
and also has the added advantage of allowing for observational errors to be accounted for.
While one might a priori expect that gas discs would be thin (because collisions between
gas clouds would cause them to quickly settle down to a thin disc), we find that
the HI discs of faint dwarf irregulars are quite thick, with mean axial ratio 
$<q> \sim 0.6$. While this is substantially larger than the typical value of $\sim 0.2$
for the {\it stellar} discs of large spiral galaxies, it is consistent with the
much larger ratio of velocity dispersion to rotational velocity ($\sigma/v_c$) in
dwarf galaxy HI discs as compared to that in spiral galaxies. Our findings have implications
for studies of the mass distribution and the Tully - Fisher relation for faint
dwarf irregular galaxies, where it is often assumed that the gas is in a thin
disc.

\end{abstract}

\begin{keywords}
galaxies: dwarf -- galaxies: irregular --  radio lines: galaxies
\end{keywords}

\section{Introduction}
\label{sec:int}

     The intrinsic shape of galaxies is interesting from a variety of perspectives.
For a given galaxy the  shape should be consistent with the dynamical model of the galaxy, while, 
for a sample of galaxies, one would expect that a correct evolutionary model should be able
to reproduce the observed distribution of shapes. It is generally assumed that disc galaxies 
can be approximated to be oblate spheroids (e.g. \cite{hubble26,sandage70,ryden06}). If one 
further assumes that the galaxies have a well defined mean axial ratio (q$_0$), then the 
observed axial ratio can be used to determine the inclination of the disc. In turn, the  
inclination is a crucial input in dynamical modeling (e.g. for the mass and structure of 
the dark matter halo), studying the Tully-Fisher relation etc. 

   The observed shape of a galaxy differs from the intrinsic shape because of projection
effects. If one has a sample of galaxies drawn from a population with a well defined intrinsic
axial ratio distribution and with random orientations with respect to the
earth, then one can determine the distribution of intrinsic axial ratios from the
observed axial ratio distribution \citep[for e.g.~][]{noe79,bin81,lam92,ryden06}. It is 
worth noting that most of these studies have focused on large galaxies, and that there 
have been relatively few that focused on dwarfs. For bright spiral galaxies, q$_0$ is often taken 
to be $\sim 0.2$ \citep[see e.g.~][]{haynes84,verheijen01}. It has also long been appreciated 
that the axial ratio is a function of Hubble type. For example, \cite{heidmann72} found 
that while discs get thinner as one goes from galaxies of morphological type Sa to Sd, 
there is a rapid increase in disc thickness as one goes from Sd to dwarf Irregular
galaxies.
Similarly, \cite{sta92} found that dwarf galaxies from the UGC catalog have 
q$_0 \sim 0.5$. 
Axial ratio is also a function of the wavelength of observation. 
For example, \citet{ryden06} showed that older populations as traced by redder stars have thicker ratios than the corresponding B band disc.
But all of these studies refer to the {\it stellar} discs of the galaxies.
Due to collisions between gas clouds, one would 
expect that the gas discs would be intrinsically quite thin, for example, for our Galaxy,
the scale height of the middle disc is only $\sim 300$~pc. In extremely gas rich dwarf galaxies
one might then expect that the gas discs are relatively thin, even though the stellar
disc may have a large axial ratio.

  We present here a study of the axial ratio distribution of the {\it HI discs} of extremely
faint dwarf irregular galaxies. We select a sample of galaxies for which HI synthesis
observations are available from the Faint Irregular Galaxy GMRT Survey (FIGGS, \cite{ay08}). The 
sample selection and analysis methods used are presented in section~\ref{sec:sample}, while
the results are discussed in section~\ref{sec:dis}. To the best of our knowledge, this
is the first such study of the intrinsic shape of the HI discs of galaxies.

\section{Sample Selection, Analysis and Results}
\label{sec:sample}
\begin{figure}
\psfig{file=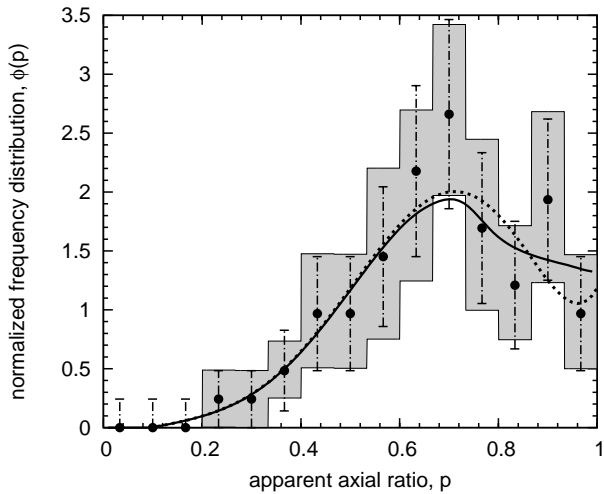,height=3.4truein, angle=270}
\caption{Histograms of the measured apparent axial ratios (after correction for the finite 
beam size). There are a total of 15 bins, the error bars were determined assuming Poisson
statistics. The shaded area represents the 80\% confidence limit histograms generated by 
bootstrap re-sampling of the data (see text for details). The dashed line shows the 
best fit (i.e. using the weighted least square) polynomial to the binned data. The solid
black line is the reconstructed distribution ($\Phi(p)$) given by  Lucy deconvolution 
(see text for details). All values have been normalized so that the area under the curves
and histogram is unity.}
\label{fig:bind}
\end{figure}

      The FIGGS sample was selected to satisfy the following criteria: (1) absolute blue magnitude
M$_{\rm B} > -14.5$, (2) integrated HI flux $> 1$ Jy~km/s and (3)~optical major axis $> 1^{'}$.
There are total of 62 galaxies in the FIGGS sample, for all of which we have axis ratios available.
We discuss the implications of the selection criteria on our derived axis ratio in 
Sec.~\ref{sec:dis}; here we merely note that any bias in the sample is likely to be
towards having a preference for edge on systems. The FIGGS survey presents HI images at a 
range of angular resolutions. Since we are interested in the overall shape of the disc,
we measured the axis ratio of the disc using the coarsest resolution HI maps.
This was done because we wanted to avoid the fine structure which shows up at higher resolutions,  and instead trace the complete spatial spread of HI which requires observing the less dense gas using coarser resolution.
 For all
of these galaxies, elliptical isophotes were fit to the HI image using the {\small STSDAS}
package in {\small IRAF}. The axis ratio that we use in this paper was measured for the
isophote with mean column density $10^{19}$ cm$^{-2}$ for all galaxies. The measured
axial ratio was corrected for the finite beam size before being used.

      For a sample of randomly oriented axi-symmetric oblate spheroids, with intrinsic axial ratio
distribution $\psi(q)$, where the axial ratio $q$ is defined as $q = c/a$, with $c$ being 
the short axis (``thickness'')  of the spheroid and $a$ the long axis (``diameter''), the 
observed distribution of axis ratios $\phi(p)$, where $p$ is the observed axial ratio
is given by (see e.g. \cite{bin78})

\begin{equation}
{\phi(p)~=~p{\displaystyle\int_{0}^{p}}{\frac{\psi(q)dq}{\sqrt{(1-q^2)(p^2-q^2)}}}}
\label{eqn:i1}
\end{equation}

   What is available to us is \phip~ the observed axial ratio distribution; in order to 
derive the intrinsic axial ratio distribution \psiq~, we have to invert Eqn.~\ref{eqn:i1}.

\begin{figure}
\psfig{file=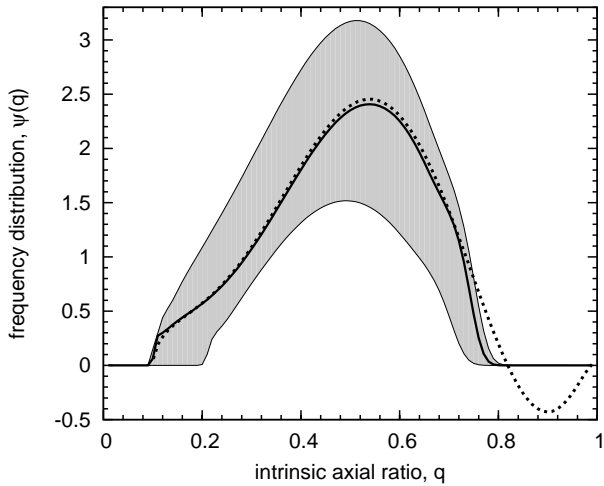,height=3.4truein, angle=270}
\caption{The frequency distribution of {\it intrinsic} axial ratios. The dashed line indicates 
\psiq~ as obtained from the direct inversion of the polynomial fit to the binned data. The solid
line is the final \psiq~  obtained after Lucy deconvolution (see text for details). The shaded 
area represents the region between the \psiq s obtained by applying Lucy deconvolution to the 
polynomial fits to the 80\% confidence interval histograms (see text for details).
}
\label{fig:psi}
\end{figure}

\citet{fal83} provide an integration formula (Eqn.~(7) in their paper) in the case that 
\phip~ is a polynomial function. This allows for a direct inversion of Eqn.~\ref{eqn:i1}. The 
binned data (Fig.~\ref{fig:bind}) were hence fit by a polynomial, using a weighted least 
squares fit. The errors were assumed to be Poisson distributed (with the error for bins with 
zero counts taken to be that for a single count). The errors were also estimated using bootstrap 
resampling. 10000 different realizations of the data were constructed using boot strap resampling,
and the 80\% confidence limit histograms determined in this way are also shown in Fig.~\ref{fig:bind}.
As can be seen, the two error estimates are in good agreement. The direct inversion of the best 
fit polynomial is shown in Fig.~\ref{fig:psi}. It has a broad peak around q$_0 \sim 0.6$, but 
also falls to negative values beyond  $q~\sim~0.8$, which is unphysical. 

While there may exist an acceptable polynomial fit to \phip~ which does not produce negatives 
on inversion, we do not investigate this possibility directly. Instead we choose to try to invert
Eqn.~\ref{eqn:i1} using Lucy's deconvolution algorithm \citep{luc74}. This approach  has the added 
advantage of also being able to account for measurement errors. The Lucy deconvolution
method is  an iterative algorithm for estimating the intrinsic frequency distribution from an observed 
distribution, subject to the constraints that the deduced distribution should be normalized as well 
as positive for all values of the intrinsic quantity ($q$ in our case). One starts with an initial guess
initial guess for the intrinsic distribution (\psiq), and uses a kernel ($k(p|q)$, defined by 
Eqn.~\ref{eqn:a4} below) to iteratively find better approximations to both the apparent distribution 
and the intrinsic distribution. 

   If we rewrite Eqn.~\ref{eqn:i1} as:
\begin{equation}
\phi(p) = {\displaystyle\int}\psi(q)k(p|q)dq
\label{eqn:a4}
\end{equation}
\noindent
then the $r$th loop of the iterative algorithm will be,
\begin{equation}
\Phi^r(p)~=~{\displaystyle\int}\psi^r(q)k(p|q)dq
\label{eqn:a5}
\end{equation}
\begin{equation}
\psi^{r+1}(q)~=~\psi^r(q){\displaystyle\int}{\frac{\phi(p)}{\Phi^r(p)}}k(p|q)dp
\label{eqn:a6}
\end{equation}

Eqn.~\ref{eqn:a6} neglects the measurement errors in determining the observed axial ratio.
Following \citet{bin81}, we can account for these errors by assuming the following form for the 
probability $E(p|p')\delta p$, that a galaxy of actual apparent axial ratio $p'$ is instead
measured to have an axial ratio $p$:

\begin{equation}
E(p|p') = {\frac{1}{\sqrt{2\pi}\sigma}}[\exp({\frac{-(p-p')^2}{2\sigma^2}})+
                                        \exp({\frac{-(2-(p+p'))^2}{2\sigma^2}})]
\label{eqn:a7}
\end{equation}

In our case, we estimate the measurement errors by fitting a straight line to the distribution of errors 
in the axial ratio $p$ as reported by the isophote fitting package. This gives:

\begin{equation}
\sigma(p')~=~0.029p'~+~0.0006
\label{eqn:a8}
\end{equation}
\noindent

and projection kernel hence becomes,

\begin{equation}
K(p|q) = {\displaystyle\int_{q}^{1}}E(p|p')k(p'|q)dp'
\label{eqn:a9}
\end{equation}

The algorithm now involves using $K(p|q)$ instead of $k(p|q)$ in equations~\ref{eqn:a5} 
and \ref{eqn:a6}, for which the range of integration now becomes 0 to 1, since now 
$K(p|q)\not=0$ for $p<q$. One would expect that in successive iterations, the approximation 
$\Phi(p)$ to the observed distribution \phip~ improve, and this can be used to decide when 
to terminate the algorithm.

Two different initial guesses for \psiq~($\psi^0(q)$) were tried, the first was peaked around 
$q \sim 0.6$, the second was constant for all $q$. These two initial guesses thus cover two
extremes of possible \psiq. For both initial guesses, \psiq~ quickly converges to the form 
shown in Fig.~\ref{fig:psi}. To determine the indicative errors in the determination of 
\psiq~, the 80\% confidence limit histograms shown in Fig.~\ref{fig:bind} were fit with 
polynomials and Lucy deconvolution was applied to these polynomials. The resulting intrinsic 
distribution functions are also shown in Fig.~\ref{fig:psi}, and the shaded area hence represents the
$\sim$ 80\% confidence interval for \psiq. The final $\Phi(p)$ (i.e. the final approximation to the
observed distribution function) is shown in Fig.~\ref{fig:bind}, and as it can be seen, it fits with the
observed distribution within the error bars. The \psiq~ shown in Fig.~\ref{fig:psi} has $<q>~=~0.57$, standard deviation $\sigma~=~0.164$,  
and skewness,~$\gamma_1~=~q^3/\sigma^3~=~-0.62$.

\section{Discussion}
\label{sec:dis}

     Eqn.~\ref{eqn:i1} (and hence results obtained from inverting it also) applies only for a sample
of randomly oriented oblate spheroids. The basic assumption hence is that the galaxy sample we are
working with has an unbiased distribution of inclination angles. The selection criteria for the
FIGGS sample (from which the current sample is drawn) includes a requirement that the optical
major axis of the galaxy be  $> 1^{'}$. Dwarf galaxies are generally dust poor \citep[for eg. see ][]{wal07,gal09}, and hence to a good
approximation have optically thin discs. Highly inclined galaxies will hence be over represented
in a diameter limited sample, i.e. our sample is biased towards edge on discs. This means that 
the true mean intrinsic axial ratio would be even larger than what we have estimated above.
It is worth noting that as the intrinsic axial ratio gets closer to 1.0, the magnitude of this 
bias decreases, and hence the bias in our estimate should not be large.

     The mean intrinsic axial ratio that we obtain, viz. $<q> \sim 0.57$ is substantially larger
than the value of $0.2$ usually adopted for the {\it stellar } discs of large spiral galaxies.
The value from our sample is more than twice as large as older measurements of $<q>$ in stellar discs of Magellanic irregular galaxies by \citet{heidmann72} (ranging from 0.20 to 0.24).
Consistent with this, our sample contains no very flat galaxies. In fact, as can be seen from
Fig.~\ref{fig:bind} there are no galaxies with apparent axial ratio $<0.2$ in the sample. 
Further, if we look at the distribution of the observed axial ratio of different classes of galaxies 
in the Automated Photographic Measuring (APM) survey as in \citet{lam92}, then our histogram resembles those for ellipticals and S0s 
more closely than that for spirals. This is another qualitative indication that the underlying 
intrinsic distribution of axial ratios has a higher mean than is typical for spirals. Interestingly,
the value of $<q>$ we obtain matches well with what \citet{sta92,binggeli95} derived for the {\it stellar} 
discs in dwarfs.

The thickness of the {\it gas } discs of dwarf galaxies is contrary to what one might have naively expected for a gas disc, since,
in general, collisions between gas clouds should cause them to quickly settle into a thin disc. 
However, this large axial ratio is probably consistent with the large gas dispersion in comparison 
to the rotational velocity observed in dwarf galaxies.
For example, \citet{kau07} did particle hydrodynamic simulations to show that dwarf galaxies with rotational velocities $\sim$40 \kms~did not originate as thin discs but thick systems. 
This still leaves open the question
of where the large velocity dispersion comes from. \cite{dut09} find good evidence for
a scale free power spectrum of HI fluctuations in dwarf galaxies, consisent with what would
be expected from a turbulent medium. Interestingly, they also find that that the dwarfs must 
have relatively thick gas discs, similar to the conclusions reached here. Assuming that
the origin of the velocity dispersion is turbulent motions in the ISM, the timescale for
dissipation is given by $\tau \sim L/v_{turb} \sim L/\sigma$ \citep[e.g.]{shu87}. The total 
turbulent energy is $E_{turb} \sim 1/2 M_{HI} \sigma^2$. For our sample galaxies, the typical HI mass
is $\sim 3 \times 10^7$~M$_\odot$, while the length scale $L \sim 1$kpc. The rate of
turbulent energy dissipation is hence $\sim 10^{44}$erg/yr. On the other hand, the 
star formation rate is $\sim 10^{-3}$ M$_\odot$/yr \citep{roychowdhury09}, for which 
the expected supernova rate for a Salpeter IMF is $\sim 7 \times 10^{-6}$/yr \citep{bin01}. 
Assuming that each supernova explosion deposits $\sim 10^{51}$ ergs of energy into 
the ISM, the energy input from star formation is $\sim 10^{46}$ ergs/yr, more than
sufficient to balance the turbulent energy loss. Thus, star formation driven turbulence
in the ISM is a plausible cause for the thick gas discs that we observe.

\begin{figure}
\psfig{file=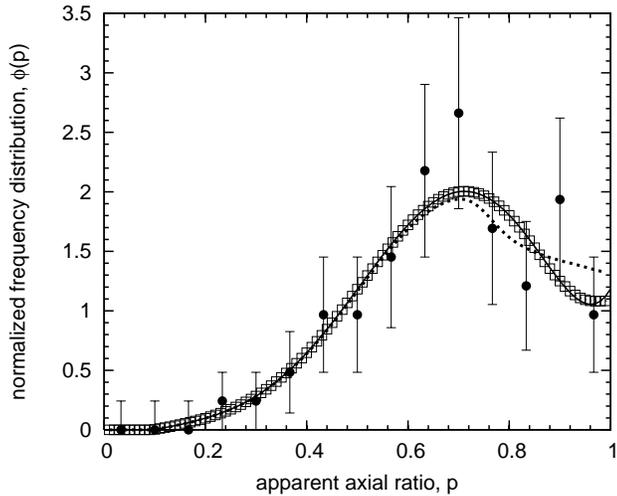,height=3.4truein, angle=270}
\caption{Reconstructed axial ratio distribution ($\Phi(p)$) assuming the gas discs to  be prolate spheroids, is shown as boxes, which follows the best fit polynomial to the binned data (solid line) closely.
Binned data with errorbars and the reconstructed distribution $\Phi(p)$ obtained using the oblate spheroid assumption (dashed line) is shown for comparison. All values have been normalized to unity.}
\label{fig:prophi}
\end{figure}

\begin{figure}
\psfig{file=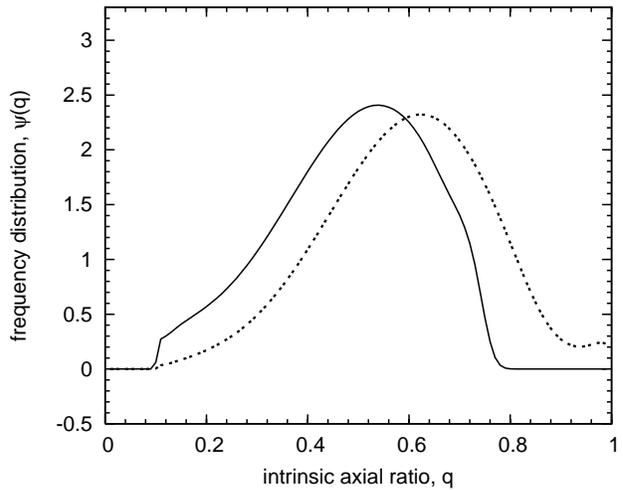,height=3.4truein, angle=270}
\caption{The frequency distribution of {\it intrinsic} axial ratios obtained after Lucy deconvolution.
The dashed line indicates \psiq~ obtained assuming the gas discs to be prolate spheroids, while the solid line shows the earlier \psiq~ obtained with the oblate spheroid assumption.
}
\label{fig:propsi}
\end{figure}

    We have assumed through out that the HI discs of dwarf galaxies can be approximated as oblate
spheroids. On the other hand, in Sec.~\ref{sec:sample} we saw that the inversion based on the
best fit polynomial to the observed histogram of axial ratios actually gives unphysical results, i.e. 
that the derived intrinsic axial ratio distribution \psiq\ becomes negative. \cite{lam92} found
a similar pattern for the axial ratio distribution of spiral and S0 galaxies in the APM catalog,
and hence relaxed the assumption that the discs are oblate spheroids. Adequate fits to their
data could be obtained assuming that the galaxies have a triaxial shape. Similarly, \cite{ryden06}
from a study of large galaxies in the 2MASS catalog concluded that spiral galaxies are mildly
triaxial. The fact that the polynomial approximation gives unphysical results for our sample
also suggests that triaxial models may provide a better fit. On the other hand, the Lucy 
deconvolution gives a physically plausible (as indeed it is designed to) intrinsic axial 
ratio distribution, that fits the observed data within the error bars. It is worth noting 
that distribution found by Lucy  deconvolution  is in excellent agreement with that found 
by direct inversion of the polynomial fit for the entire range for which the latter is $>0$. Interestingly, if we assume the gas discs to be prolate instead of oblate spheroids, Lucy deconvolution produces an equally acceptable \phip~ as can be seen from Fig.~\ref{fig:prophi}.
The \psiq~ obtained from Lucy deconvolution with a prolate spheroid assumption (see Fig.~\ref{fig:propsi}) indicates that such gas discs should be more cylindrical than spherical. However, the observed
kinematics of gas in these galaxies \citep[see ][etc.]{bch03,bc03,bc04a,bc04b,bck05,bcks05,beg06,ay08} shows that the gas has a significant rotational support. As discussed earlier, a gravitationally bound rotating disc of gas forms an oblate and not a prolate spheroid. Thus, although from the axial ratio data alone,
one cannot distinguish between a prolate and oblate shapes, in
conjuction with the kinematical data, it is clear that the gas
is distributed in the form of a thick disc.

    In summary, we find that the {\it gas} discs of dwarf galaxies are relatively thick,
in sharp contrast to the gas discs of spiral galaxies. This has implications both for
the internal dynamics of the gas, as well as for studies of the mass distribution 
Tully - Fisher relation in faint dwarfs,  in which it is generally assumed that the gas
is in a thin disc.

\section*{Acknowledgments}
We thank the staff of the GMRT who have made the observations used in this paper possible.
GMRT is run by the National Centre for Radio Astrophysics of the Tata Institute of Fundamental Research.

\bsp

\label{lastpage}


\begin{thebibliography}{}

\bibitem[\protect\citeauthoryear{Begum et al.}{2003}]{bch03} Begum, A.,  
	Chengalur, J.N. \& Hopp, U., 2003, New Astronomy, 8, 267
\bibitem[\protect\citeauthoryear{Begum \& Chengalur}{2003}]{bc03} Begum, A \& 
	Chengalur, J.N., 2003, A\&A, 409, 879
\bibitem[\protect\citeauthoryear{Begum \& Chengalur}{2004a}]{bc04a} Begum, A \& 
	Chengalur, J.N., 2004, A\&A, 413, 525
\bibitem[\protect\citeauthoryear{Begum \& Chengalur}{2004b}]{bc04b} Begum, A \& 
	Chengalur, J.N., 2004, A\&A, 424, 509
\bibitem[\protect\citeauthoryear{Begum, Chengalur \& Karachentsev}{2005}]{bck05} Begum, A, 
	Chengalur, J.N. \& Karachentsev, I. D., 2005, A\&A, 433, 1L
\bibitem[\protect\citeauthoryear{Begum et al.}{2005}]{bcks05} Begum, A, 
	Chengalur, J.N., Karachentsev, I. D. \& Sharina, M. E., 2005, MNRAS, 359, 53L
\bibitem[\protect\citeauthoryear{Begum et al.}{2006}]{beg06} Begum A.,
         Chengalur J. N., Karachentsev I. D., Kaisin S. S. \&  Sharina M. E., 2006, MNRAS, 365,1220
\bibitem[\protect\citeauthoryear{Begum et al.}{2008}]{ay08} Begum A.,
        Chengalur J. N., Karachentsev I. D., Sharina M. E. \& Kaisin S. S., 2008, MNRAS,
        386, 1667B
\bibitem[\protect\citeauthoryear{Binggeli \& Popescu}{1995}]{binggeli95} Binggeli B., 
        Popescu C.~C., 1995, A\&A, 298, 63
\bibitem[\protect\citeauthoryear{Binney}{2001}]{bin01} Binney J., 2001, in Hibbard J. E., Rupen M. P., von Gorkom H., eds, ASP Conf. Ser. Vol. 240, Gas and Galaxy Evolution, Astron. Soc. Pac., San Francisco, p. 355
\bibitem[\protect\citeauthoryear{Binney}{1978}]{bin78} Binney J., 1978, MNRAS, 183, 501
\bibitem[\protect\citeauthoryear{Binney \& de Vaucouleurs}{1981}]{bin81} Binney J. \& 
         de Vaucouleurs G., 1981, MNRAS, 194, 679
\bibitem[\protect\citeauthoryear{Dutta et al.}{2009}]{dut09} Dutta Prasun, Begum Ayesha, Bharadwaj Somnath, \& Chengalur Jayaram N., 2009, MNRAS, 398, 887
 \bibitem[\protect\citeauthoryear{Fall \& Frenk}{1983}]{fal83} Fall S. M. \& Frenk C. S., 1983, 
         AJ, 88(11), 1626
 \bibitem[\protect\citeauthoryear{Galametz et al.}{2009}]{gal09} Galametz M., et al., 2009, A\&A, 508, 645
\bibitem[\protect\citeauthoryear{Haynes \& Giovanelli}{1984}]{haynes84} Haynes M.~P., 
        Giovanelli R., 1984, AJ, 89, 758 
\bibitem[\protect\citeauthoryear{Heidmann, Heidmann, \& de Vaucouleurs}{1972}]{heidmann72} 
        Heidmann J., Heidmann N., de Vaucouleurs G., 1972, Mem. Roy. Astron. Soc., 75, 85
\bibitem[\protect\citeauthoryear{Hubble}{1926}]{hubble26} Hubble E.~P., 1926, ApJ, 64, 321
 \bibitem[\protect\citeauthoryear{Kaufmann, Wheeler and Bullock}{2007}]{kau07} Kaufmann T., Wheeler C., Bullock J. S., 2007, MNRAS, 382, 1187
\bibitem[\protect\citeauthoryear{Lambas, Maddox \& Loveday}{1992}]{lam92} Lambas D. G., 
        Maddox S. J. \& Loveday J., 1992, MNRAS, 258,404
\bibitem[\protect\citeauthoryear{Lucy}{1974}]{luc74} Lucy L. B., 1974, AJ, 79(6), 745
\bibitem[\protect\citeauthoryear{Noerdlinger}{1979}]{noe79} Noerdlinger Peter D., 1979, ApJ, 234, 802
\bibitem[\protect\citeauthoryear{Roychowdhury et al.}{2009}]{roychowdhury09} Roychowdhury S., 
        Chengalur J.~N., Begum A., Karachentsev I.~D., 2009, MNRAS, 397, 1435
\bibitem[\protect\citeauthoryear{Ryden}{2006}]{ryden06} Ryden B.~S., 2006, ApJ, 641, 773 
\bibitem[\protect\citeauthoryear{Sandage, Freeman, \& Stokes}{1970}]{sandage70} Sandage A., 
         Freeman K.~C., Stokes N.~R., 1970, ApJ, 160, 831 
\bibitem[\protect\citeauthoryear{Shu, Adams, \& Lizano}{1987}]{shu87} Shu F.~H., Adams F.~C., 
         Lizano S., 1987, ARA\&A, 25, 23
\bibitem[\protect\citeauthoryear{Staveley-Smith, Davies \& Kinman}{1992}]{sta92} Staveley-Smith L.,
        Davies R. D. \& Kinman T.D., 1992, MNRAS, 258, 334
\bibitem[\protect\citeauthoryear{Verheijen \& Sancisi}{2001}]{verheijen01} Verheijen M.~A.~W., 
        Sancisi R., 2001, A\&A, 370, 765 
\bibitem[\protect\citeauthoryear{Walter et al.}{2007}]{wal07} Walter Fabian et al., 2007, ApJ, 661, 102

\end{thebibliography}
\end{document}